\documentclass [6pt,a4paper]{article}
\usepackage{bbm}
\usepackage{amsfonts}
\usepackage{mathrsfs}
\usepackage {amssymb}
\usepackage {amsmath}
\usepackage{amsthm}
\usepackage{latexsym}
\usepackage{booktabs}
\def\dse#1{\vskip 0.6cm\noindent
        {\large\bf #1}
        \vskip 0.4cm}

\def\dse#1{\vskip 0.6cm\noindent
        {\large\bf #1}
        \vskip 0.4cm}
\usepackage{lastpage}
\usepackage{epsfig}

\begin{document}
\begin{center}
\textbf{\large{Cyclic DNA codes over
$\mathbb{F}_2+u\mathbb{F}_2+v\mathbb{F}_2+uv\mathbb{F}_2$}}\footnote { E-mail
addresses:
 zhushixin@hfut.edu.cn(S.Zhu), chenxiaojing0909@126.com(X.Chen).\\
This research is supported by the National Natural Science
Foundation of China (No.61370089) and  the Anhui Provincial Natural Science Foundation under Grant JZ2015AKZR0229.}
\end{center}

\begin{center}
{ { Shixin Zhu, \  Xiaojing Chen} }
\end{center}

\begin{center}
\textit{\footnotesize Department of Mathematics, Hefei University of
Technology, Hefei 230009, Anhui, P.R.China \\
  }
\end{center}

\noindent\textbf{Abstract:}  In this work, we study the structure of cyclic DNA codes of arbitrary lengths over the ring
$R=\mathbb{F}_2+u\mathbb{F}_2+v\mathbb{F}_2+uv\mathbb{F}_2$ and establish relations to codes over
$R_{1}=\mathbb{F}_2+u\mathbb{F}_2$  by defining a Gray map between $R$ and $R^{2}_{1}$  where $R_{1}$ is the ring with $4$ elements. Cyclic codes of arbitrary lengths over $R$ satisfied the reverse constraint and the reverse-complement constraint are studied in this paper. The $GC$ content constraint is considered in the last.\\


\dse{1~~Introduction}
Algebraic coding theory of linear codes has attracted remarkable attention for the last half of the century(e.g. see [7,10]). Cyclic codes are important families of linear codes because of their rich algebraic structures and practical implementations. The focus on constructing codes was mainly over fields, but after the study in [11] finite rings have received a great deal of attention. Most of the studies are concentrated on codes over finite chain rings [8]. However, optimal codes over non-chain rings exist(e.g see [19]). But the case over a non-chain structure is more complicated [4]. In [22], the algebraic structure of cyclic codes over $\mathbb{F}_{2}+v\mathbb{F}_{2}$, where $v^{2}=v$ are studied. Zhu and Wang studied a class of constacyclic codes over $\mathbb{F}_{p}+v\mathbb{F}_{p}$ in [21].

On the other hand, Adleman [2] pioneered the studies on DNA computing by solving an instance of NP-complete problem over DNA molecules. DNA is a nucleic acid containing the genetic instructions used in the development and functioning of all known living organisms. It is formed by strands linked together and twisted in the shape of a double helix. Each strand is a sequence consists of four possible nucleotides, two purines, adenine $(A)$  and guanine $(G)$, and two pyrimidines, thymine $(T)$ and cytosine $(C)$. The ends of a DNA strand are chemically polar with $5^{'}$  and  $3^{'}$  ends, which implies that the strands are oriented. DNA has two strands that are governed by the rule called Watson Crick complement(WCC), that is, $A$  pairs with $T$ and $G$ pairs with $C$. We denote the WCC in this paper as $\overline{A}=T$, $\overline{T}=A$, $\overline{G}=C$ and $\overline{C}=G$. The pairing is done in the opposite direction and reverse order. For instance, the WCC strand of $3^{'}-TAAGCTC-5^{'}$ is the strand $5^{'}-GAGCTTA-3^{'}$.

Furthermore, since DNA computing can store more memory than silicon based computing systems, there are many scholars begin to study it. Siap et al. [18] constructed cyclic DNA codes considering the $GC$ content constraint over $\mathbb{F}_{2}[u]/(u^{2}-1)$ and used the deletion distance. Guenda and Gulliver [9] studied cyclic codes over $\mathbb{F}_{2}[u]/(u^{2})$ satisfy the reverse constraint and the reverse-complement constraint and the $GC$ content constraint, and an infinite family of BCH DNA codes are constructed. Recently, Liang and Wang [13] studied the cyclic DNA codes over $\mathbb{F}_{2}+u\mathbb{F}_{2}$. Yildiz and Siap [20] studied DNA pairs instead of single DNA bases for the first time, where $16$ elements of a ring and DNA pairs are matched and the algebraic structure of these DNA codes are studied. Later in [16], DNA pairs are matched with $\mathbb{F}_{16}$ and by introducing some special polynomials DNA codes are constructed. It is also observed that in some cases reversible codes introduced by Massey over $\mathbb{F}(q)$ are useful for constructing DNA codes in [15]. Recently, Bayram et al. [3] have considered codes over the ring $\mathbb{F}_{4}+v\mathbb{F}_{4}$. The constacyclic codes and skew constacyclic codes over the ring are studied. And they studied the structure of DNA codes over the ring and present applications to DNA codes.  However, there is not much work has been done on DNA codes over non-chain ring. And we do such work over the ring $\mathbb{F}_2+u\mathbb{F}_2+v\mathbb{F}_2+uv\mathbb{F}_2$.

In this paper, we study the structure of cyclic DNA codes of arbitrary lengths over the ring $R$. Cyclic codes over ring $R_{1}$ have been extensively studied by many authors [1,5,6,14]. The rest of the paper is organized as follows: Sect.2 includes some basic background and some basic results of cyclic codes of arbitrary lengths over $R_{1}$. In Sect.3, we study cyclic codes satisfy the reverse constraint and reverse-complement constraint over such ring, the existence and the structure of such codes are complemently determined. In Sect.4, we study the structure of DNA codes over $R$ and present applications to DNA codes where some examples of such codes are optimal. In Sect.5, we use the Gray images of the minimal generating set of $C$ to study the $GC$ content of $C$. Section $6$ concludes the paper.

\dse{2~~Preliminaries}
 Let $\mathbb{F}_{2}$ be the binary field. Throughout this paper $R$ denotes the commutative ring $\mathbb{F}_2+u\mathbb{F}_2+v\mathbb{F}_2+uv\mathbb{F}_2$ with $u^{2}=0$, $v^{2}=v$ and $uv=vu$ with characteristic $2$. Let $R_{1}$ be the finite chain ring $\mathbb{F}_2+u\mathbb{F}_2$ with $u^{2}=0$. $R$ is a semi-local ring with two maximal ideals namely $I_{u+v}$ and $I_{1+u+v}$. The quotient rings $R/I_{u+v}$ and $R/I_{1+u+v}$ are isomorphic to $\mathbb{F}_{2}$. A direct decomposition of $R$ is $R=I_{v}\oplus I_{1+v}$. We can also see that $I_{v}$ and  $I_{1+v}$  are isomorphic to  $R_{1}$. So every element $c$ in $R$ therefore can uniquely be written as $c=a+bv$, $a,b\in R_{1}$. $R$ is isomorphic to the residue ring $R_{1}[v]/\langle v^{2}-v\rangle$. Note that $I_{v}= \{av|a\in R_{1}\}$ and $I_{1+v}= \{b(1+v)|b\in R_{1}\}$.

An important property of codes over the ring $R$ is the existence of a mapping $\xi$ called the Gray map which sends linear codes over $R$ to binary linear codes. The Gray map from $R$ to $R^{2}_{1}$ is defined as
\begin{align}
\xi(a+bv)=(a,a+b).
\end{align}

One type of nontrivial automorphisms can be defined over $R$ as follows :
\[\sigma:\mathbb{F}_2+u\mathbb{F}_2+v\mathbb{F}_2+uv\mathbb{F}_2 \rightarrow\mathbb{F}_2+u\mathbb{F}_2+v\mathbb{F}_2+uv\mathbb{F}_2,\]
\begin{equation}
a+bv\rightarrow a+(1+v)b,~a,b\in\mathbb{F}_2+u\mathbb{F}_2.
\end{equation}
Let $C$ be a linear code over $R$. The following result is presented in [10], let
\begin{equation}
C_{1}=\{x+y\in R^{n}_{1}|(x+y)v+x(v+1)\in C,~for~some~x,y\in  R^{n}_{1}\},\end{equation}
\begin{equation}
C_{2}=\{x\in R^{n}_{1}|(x+y)v+x(v+1)\in C,~for~some~y\in  R^{n}_{1}\}.
\end{equation}
Note that $C_{1}$ and $C_{2}$ are linear codes over $R_{1}$. Consequently, $C=vC_{1}\oplus(1+v)C_{2}$.\\

\noindent\textbf{Corollary 2.1} \emph{(1)Let $C$ be a linear code over $R$ such that $C=vC_{1}\oplus(1+v)C_{2}$. Then, $C$
is a cyclic code if and only if $C_{1}$ and $C_{2}$ are both cyclic codes over $R_{1}$.\\
(2)If $C=vC_{1}\oplus(1+v)C_{2}$ is a cyclic code of length $n$ over $R$, then $C=(vf_{1},(1+v)f_{2})$ where
$f_{1}$ and $f_{2}$ are the generator polynomials of $C_{1}$ and $C_{2}$, respectively.}\\

Recall that the Hamming weight of a codeword $c$ is defined by $w_{H}(c)=|\{i|c_{i}\neq0\}|$, i.e., the number of the nonzero entries of $c$. The minimum Hamming weight $w_{H}(c)$ of a code $C$ is the smallest possible weight among all its nonzero codewords. The Hamming distance $d(c_{1},c_{2})$  between two codewords $c_{1}$ and $c_{2}$ is the Hamming weight of the codeword $c_{1}-c_{2}$. The minimum Hamming distance $d(C)$ of $C$ is defined as $min\{{d(c_{1},c_{2})|c_{1},c_{2}\in C,c_{1}\neq c_{2}}\}$.

A code is called a DNA codes if it satisfies some or all of the following conditions:\\
$(1)$ \emph{The Hamming constraint} For any two different codewords $c_{1},c_{2}\in C,H(c_{1},c_{2})\geq d$.\\
$(2)$ \emph{The reverse constraint} For any two codewords $c_{1},c_{2}\in C,H(c_{1},c^{r}_{2})\geq d$.\\
$(3)$ \emph{The reverse-complement constraint} For any two codewords $c_{1},c_{2}\in C,H(c_{1},c^{rc}_{2})\geq d$.\\
$(4)$ \emph{The fixed $GC$ content constraint} For any codeword $c\in C$ contains the same number of $G$

and $C$ elements.\\

The purpose of the first three constraints is to avoid undesirable hybridization between different strands. The fixed $GC$ content ensures that all codewords have similar thermodynamic characteristics, which allows parallel operations on DNA sequences.

The structure of cyclic codes of arbitrary lengths $n$ over $R_{1}$ has been extensively studied in [1], which is\\

\noindent\textbf{Theorem 2.2 } [1] \emph{ Let $C$ be a cyclic code in $R_{1,n}=R_{1}[x]/(x^{n}-1)$. Then\\
(1)	If $n$ is odd, then $R^{n}_{1}$ is a principal ideal ring and $C=(g,ua)=(g+ua)$, where $g$, $a$ are\\
binary polynomials with $a\mid g\mid (x^{n}-1)mod2$.\\
(2) If  $n$ is not odd, then\\
(2.1) $C=(g+up)$, where $g\mid (x^{n}-1)mod2$ and $(g+up)\mid(x^{n}-1)$ in $R$ and $g\mid p\widehat{g}$. Or,\\
(2.2) $C=(g+up,ua)$, where $g$, $a$ and $p$ are binary polynomials with $a\mid g\mid(x^{n}-1)mod2$,\\
$a\mid p\widehat{g}$ and $degp\leq dega$.}\\

\noindent\textbf{Remark 1.} In this paper, we use $f$, $\widehat{f}$ to represent $f(x)$ and $(x^{n}-1)/f(x)$ respectively if don't confuse.\\

\dse{3~~ The reverse constraint and reverse-complement constraint codes }  In this section, we main study the reverse constraint and the reverse-complement constraint codes over $R$. We begin with the following definition. For each codeword $x=(x_{0},x_{1},\cdots,x_{n-1})\in R$, we define the reverse of $x$ as  $x^{r}=(x_{n-1},x_{n-2},\cdots,x_{0})$, the complement of  $x$ as $x^{c}=(\overline{x_{0}},\overline{x_{1}},\cdots,\overline{x_{n-1}})$ and the reverse-complement of $x$ as $x^{rc}=(\overline{x_{n-1}},\overline{x_{n-2}},\cdots,\overline{x_{0}})$. Furthermore, for each polynomial $c(x)=c_{0}+c_{1}x+\cdots+c_{r}x^{r}$ with $c_{r}\neq0$, the reciprocal of $c(x)$ is defined to be the polynomial $c^{*}(x)=x^{r}c(x^{-1})=c_{r}+c_{r-1}x+\cdots+c_{0}x^{r}$. We note that $degc^{*}(x)\leq degc(x)$ and if $c_{0}\neq0$, then $c(x)$ and $c^{*}(x)$ always have the same degrees. $c(x)$ is called self-reciprocal if and only if $c(x)$=$c^{*}(x)$.\\

Let $S_{D_{4}}={A,T,C,G}$ represent the DNA alphabet. We use the same notation for the set
\begin{equation}
S_{D_{16}}=\{AA,AT,AC,AG,TT,TA,TC,TG,CC,CA,CT,CG,GG,GA,GT,GC\},\end{equation}
which is originally presented in [16]. We define a $\zeta$ correspondence between the elements of the ring $R$ and DNA double pairs presented explicitely in Table 1. The elements $0,1,u,1+u$ of $R_{1}$ are in one-to-one correspondence with the nucleotide DNA bases $A,T,C,G$ such that $0\rightarrow A$, $1\rightarrow G$, $u\rightarrow T$ and $1+u\rightarrow C$. The Watson Crick complement is given by $\overline{A}=T$, $\overline{T}=A$, $\overline{G}=C$ and $\overline{C}=G$. Naturally we extend this notion to the elements of $S_{D_{16}}$ such that $\overline{AA}=TT,\cdots,\overline{TG}=AC$.\\

\noindent\textbf{Definition 3.1}  Let $C$ be a code over $R$ of arbitrary lengths $n$ and $c\in C$ be a codeword where $c=(c_{0},c_{1},\cdots,c_{n-1})$, $c_{i}\in R$, then we define
\[\Phi (c):C\rightarrow S^{2n}_{D_{4}},\]
\begin{equation}
(a_{0}+b_{0}v,a_{1}+b_{1}v,\cdots,a_{n-1}+b_{n-1}v)\mapsto (a_{0},a_{1},\cdots,a_{n-1},a_{0}+b_{0},a_{1}+b_{1},\cdots,a_{n-1}+b_{n-1}).
\end{equation}
by using Table 1.\\

\begin{table}[htbp]
{\small
\begin{center}
{\small{\bf Table 1 }~~$\zeta-$table for DNA correspondence}\\
\begin{tabular}{c c c}\\
\hline
Elements a & Gray images & DNA double pairs $\zeta(a)$ \\
\hline
$0$ & $(0,0)$ & $AA$ \\
 $v$ & $(0,1)$ & $AG$ \\
 $uv$ & $(0,u)$ & $AT$ \\
 $v+uv$ & $(0,1+u)$ & $AC$ \\
 $1$ & $(1,1)$ & $GG$ \\
 $1+v$ & $(1,0)$ & $GA$ \\
 $1+uv$ & $(1,u+1)$ & $GC$ \\
 $1+v+uv$ & $(1,u)$ & $GT$ \\
 $u$ & $(u,u)$ & $TT$ \\
 $u+v$ & $(u,1+u)$ & $TC$ \\
 $u+uv$ & $(u,0)$ & $TA$ \\
 $u+v+uv$ & $(u,1)$ & $TG$ \\
 $1+u$ & $(1+u,1+u)$ & $CC$ \\
 $1+u+v$ & $(1+u,u)$ & $CT$ \\
 $1+u+uv$ & $(1+u,1)$ & $CG$ \\
 $1+u+v+uv$ & $(1+u,0)$ & $CA$ \\
\hline
\end{tabular}
\end{center}}
\end{table}

For instance, $(c_{0},c_{1},c_{2},c_{3})=(1,v,u,u+v)$ is mapped to
\begin{equation}
\Phi(1,v,u,u+v)=(GATTGGTC).
\end{equation}

\noindent\textbf{Definition 3.2} A cyclic code $C$ of length $n$ over $R$ is said to be reversible if $x^{r}\in C$ for all $x\in C$, complement if $x^{c}\in C$ for all $x\in C$ and reversible-complement if $x^{rc}\in C$ for all $x\in C$.
\\

\noindent\textbf{Lemma 3.3} \emph{Let $f$, $g$ be any two polynomials in $R$ with $degg\leq degf$. Then
\begin{equation}
1.~(f\cdot g)^{*}=f^{*}\cdot g^{*};\end{equation}
\begin{equation}
2.~(f+g)^{*}=f^{*}+x^{degf-degg}g^{*}.\end{equation}}

\dse{3.1~~The reverse constraint codes } The following result is due to Massey [15, Theorem 1]. It characterizes the reversible codes over finite fields.\\

\noindent\textbf{Lemma 3.4} [15] \emph{Let $C=(f)$ be a cyclic code over $\mathbb{F}_{2}$ where $f$ is a monic polynomial, then $C$ is reversible if and only if $f$ is self-reciprocal.}\\

The reverse constraint on cyclic codes of arbitrary lengths over $R_{1}$ has been studied in [9] and [13], we list it for convenient in our later study.\\

\noindent\textbf{Lemma 3.5} [9] \emph{Let $C=(g,ua)=(g+ua)$ be a cyclic code of odd length $n$ over $R_{1}$. Then $C$ is reversible if and only if $g$ and $a$ are self-reciprocal.}\\

\noindent\textbf{Lemma 3.6} [13] \emph{Let $C=(g+up)$ be a cyclic code of even length $n$ over $R_{1}$. Then $C$ is reversible if and only if\\
1.~$g$ is self-reciprocal;\\
2.~(a) $x^{i}p^{*}=p$. Or\\
(b) $g=x^{i}p^{*}+p$, where $i=degg-degp$.}\\

\noindent\textbf{Lemma 3.7} [13] \emph{Let $C=(g+up,ua)$ with $a\mid g\mid(x^{n}-1)mod2$, $a\mid p\widehat{g}$ and $degp\leq dega$ be a cyclic code of even length $n$ over $R_{1}$. Then $C$ is reversible if and only if\\
1. $g$ and $a$ are self-reciprocal;\\
2. $a\mid(x^{i}p^{*}+p)$, where $i=degg-degp$.}\\

We will give one of the main conclutions below.\\

\noindent\textbf{Theorem 3.8} \emph{Let $C=vC_{1}\oplus(1+v)C_{2}$ be a cyclic code of arbitrary lengths $n$ over $R$. Then $C$ is reversible if and only if $C_{1}$ and $C_{2}$ are reversible, respectively, where $C_{1}$ and $C_{2}$ are both cyclic codes over $R_{1}$.}\\

\noindent\textbf{Proof.} If $C_{1}$ and $C_{2}$ are reversible, we have $C^{r}_{1}\in C$ and $C^{r}_{2}\in C$. For any  $b\in C$, $b=vb_{1}+(1+v)b_{2}$ where $b_{1}\in C_{1}$ and $b_{2}\in C_{2}$. We can easy know that $b^{r}_{1}\in C_{1}$  and $b^{r}_{2}\in C_{2}$, thus $b^{r}=vb^{r}_{1}+(1+v)b^{r}_{2}\in C$. Hence $C$ is reversible.

On the other hand, if $C$ is reversible, then for any $b=vb_{1}+(1+v)b_{2}\in C$, where $b_{1}\in C_{1},b_{2}\in C_{2}$. we have $b^{r}=vb^{r}_{1}+(1+v)b^{r}_{2}\in C$. Let $b^{r}=vb^{r}_{1}+(1+v)b^{r}_{2}=ve_{1}+(1+v)e_{2}$, where  $e_{1}\in C_{1},e_{2}\in C_{2}$. Then $v(b^{r}_{1}-e_{1})+(1+v)(b^{r}_{2}-e_{2})=0$, thus $b^{r}_{1}=e_{1}\in C_{1}$ and $b^{r}_{2}=e_{2}\in C_{2}$. Hence $C_{1}$ and $C_{2}$ are reversible, respectively.\qed\\

\noindent\textbf{Example 3.9} Let $x^{8}-1=(x+1)^{8}=g^{8}$ over $\mathbb{F}_{2}$. Let $C_{1}=(f_{1})=(g_{1}+up_{1})$, $g_{1}=g^{6}$, $p_{1}=x^{5}+x$, $C_{2}=(f_{2})=(g_{2}+up_{2})$, $g_{2}=g^{4}$, $p_{2}=x^{3}+x$. It is easy to check that $g_{1}$ and $g_{2}$ are self-reciprocal, $x^{i}p^{*}_{1}=p_{1}$ and $x^{j}p^{*}_{2}=p_{2}$, where $i=degg_{1}-degp_{1}$, $j=degg_{2}-degp_{2}$. Since $C=(f)=(v(g_{1}+up_{1})+(1+v)(g_{2}+up_{2}))$, clearly we have $f=vx^{6}+uvx^{5}+x^{4}+(u+uv)x^{3}+vx^{2}+ux+1\in C$, $f^{r}=vx+uvx^{2}+x^{3}+(u+uv)x^{4}+vx^{5}+ux^{6}+x^{7}$. On the other hand,
\begin{equation}
(vx+(1+v)x^{3})f=vx+uvx^{2}+x^{3}+(u+uv)x^{4}+vx^{5}+ux^{6}+x^{7}=f^{r}\in C.\end{equation}
By the Theorem 3.8, $C$ is a reversible code of length $8$ over $R$.\\

\dse{3.2~~The reverse-complement constraint codes } In this section, cyclic codes of arbitrary lengths satisfy the reverse-complement are examined. We give some useful lemmas firstly which can be easily proved.\\

\noindent\textbf{Lemma 3.10} \emph{For any $c\in R$, we have $c+\overline{c}=u$.}\\

\noindent\textbf{Lemma 3.11} \emph{Let $a,b\in R$, then $\overline{a+b}=\overline{a}+\overline{b}+u$.}\\

\noindent\textbf{Lemma 3.12} \emph{If $c\in \mathbb{F}_{2}$, then we have $u+\overline{uc}=uc$.}\\

We will give one of our main conclusions below.\\

\noindent\textbf{Theorem 3.13} \emph{Let $C=vC_{1}\oplus(1+v)C_{2}$ be a cyclic code of arbitrary length $n$ over $R$. Then $C$ is reversible-complement if and only if $C$ is reversible and $(\overline{0},\overline{0},\cdots,\overline{0})\in C$, where $C_{1}$ and $C_{2}$ are both cyclic codes over $R_{1}$.}\\

\noindent\textbf{Proof.} Suppose $C=vC_{1}\oplus(1+v)C_{2}$, where $C_{1}$ and $C_{2}$ are both cyclic codes over $R_{1}$. For any $c=(c_{0},c_{1},\cdots,c_{n-1})\in C$, $c^{rc}=(\overline{c_{n-1}},\overline{c_{n-2}},\cdots,\overline{c_{0}})\in C$ because of $C$ is reversible-complement. Since the zero codeword is in $C$ then its WCC is also in $C$, i.e.,
\begin{equation}
(\overline{0},\overline{0},\cdots,\overline{0})\in C\end{equation}
Whence,
\begin{equation}
c^{r}=(c_{n-1},c_{n-2},\cdots,c_{0})=(\overline{c_{n-1}},\overline{c_{n-2}},\cdots,\overline{c_{0}})+(\overline{0},\overline{0},\cdots,\overline{0})\in C.\end{equation}

On the other hand, if $C$ is reversible, then for any $c=(c_{0},c_{1},\cdots,c_{n-1})\in C$, $c^{r}=(c_{n-1},c_{n-2},\cdots,c_{0})\in C$. Since $(\overline{0},\overline{0},\cdots,\overline{0})\in C$, we get
\begin{equation}
c^{rc}=(\overline{c_{n-1}},\overline{c_{n-2}},\cdots,\overline{c_{0}})=(c_{n-1},c_{n-2},\cdots,c_{0})+(\overline{0},\overline{0},\cdots,\overline{0})\in C.\end{equation}
Therefore, $C$ is reversible-complement.\qed\\

\noindent\textbf{Example 3.14} In Example 3.9, since $C$ is reversible, if $(\overline{0},\overline{0},\cdots,\overline{0})\in C$, we can get $C$ is reversible-complement immediately.\\

Let $C$ be a cyclic code of arbitrary lengths $n$ over $R$. Then we can get the conditions that $C$ is reversible or reversible-complement easily by using Corollary 2.1, Theorem 2.2, Lemma 3.5, Lemma 3.6, Lemma 3.7, Theorem 3.8 and Theorem 3.13.\\

\dse{4 ~~DNA codes over $R$  } In this section, the design of linear DNA codes is presented. We obtain DNA codes over $R$ of arbitrary lengths that correspond to DNA double pairs.\\

\noindent\textbf{Definition 4.1} Let $f_{1}$ and $f_{2}$ be polynomials with $degf_{1}=t_{1}$, $degf_{2}=t_{2}$ and both dividing $x^{n}-1$ over $R_{1}$. Let $m=min\{n-t_{1},n-t_{2}\}$ and $f=vf_{1}+(1+v)f_{2}$ over $R$. The set $L(f)$ is called a $\sigma$-set and is defined as  $L(f)=\{E_{0},E_{1},\cdots,E_{m-1},F_{0},F_{1},\cdots,F_{m-1}\}$ where
$E(i)=x^{i}f,F_{i}=x^{i}\sigma(h)$, $0\leq i\leq m-1$, $h=vx^{t_{2}-t_{1}}f_{1}+(1+v)f_{2}$ if $t_{2}\geq t_{1}$, $h=vf_{1}+(1+v)x^{t_{1}-t_{2}}f_{2}$ otherwise.\\

$L(f)$ generates a linear code $C$ over $R$ denoted by $C=\langle f\rangle_{\sigma}$.\\

\noindent\textbf{Remark 2.} In this paper, the notation $\langle L(f)\rangle$ or $\langle f\rangle_{\sigma}$ denotes the $R$-module generated by the set $L(f)$. The notation $(f)$ stands for the ideal generated by $f$.\\

Let $f=a_{0}+a_{1}x+a_{2}+\cdots+a_{t}x^{t}$ over $R$, $\sigma(h)=b_{0}+b_{1}x+\cdots+b_{s}x^{s}$ and the $R$-submodule generated by $L(f)$ can be considered to be generated by the rows of following matrix

\begin{equation}
 L(f)=
 \begin{pmatrix}
                    E_{0}   \\
                    F_{0}  \\
                    E_{1}  \\
                    F_{1}  \\
                   \vdots
\end{pmatrix}
= \begin{pmatrix}
                     a_{0} & a_{1} & a_{2} & \cdots & a_{t} & \cdots & 0 & \cdots & \cdots & 0   \\
                     b_{0} & b_{1} & b_{2} & \cdots & \cdots  & \cdots & b_{s} & \cdots & \cdots & 0   \\
                     0 & a_{0} & a_{1} & \cdots  &\cdots & a_{t} & \cdots  & \cdots & \cdots & 0  \\
                     0 & b_{0} & b_{1} & \cdots  & \cdots & \cdots  & \cdots & b_{s}& \cdots & 0  \\
                     \cdots & \cdots & \cdots & \cdots & \cdots & \cdots & \cdots & \cdots & \cdots & \cdots
\end{pmatrix}.
\end{equation}

\noindent\textbf{Theorem 4.2} \emph{Let $f_{1}$ and $f_{2}$ be self-reciprocal polynomials dividing $x^{n}-1$ over $R_{1}$ with degree $t_{1}$ and $t_{2}$. If $f_{1}=f_{2}$, then $f=vf_{1}+(1+v)f_{2}$ and $|\langle L(f)\rangle|=16^{m}$.
Besides, $C=\langle L(f)\rangle$ is a linear code over $R$ and $\Phi(C)$ is a reversible DNA code.}\\

\noindent\textbf{Proof.} Most of the claims follow from the algebraic structures that are discussed before. Especially, the reverse of each DNA code given by $C=\langle L(f)\rangle$ over $R$ is shown to fall inside the codes by the following observation
\begin{align}
(\Phi(\sum\alpha_{i}E_{i}+\sum\beta_{i}F_{i}))^{r}=\Phi(\sum\sigma(\alpha_{i})F_{m-1-i}+\sum\sigma(\beta_{i})E_{m-1-i}),
\end{align}
where $\alpha_{i}, \beta_{i}\in R$ and $0\leq i\leq m-1$.\qed\\

Below we give an example that illustrates the power of Theorem 4.2.\\

\noindent\textbf{Example 4.3} Let $f_{1}=x+1$ and $f_{2}=x^{6}+x^{3}+1$ where both divide $x^{9}-1$ over $\mathbb{F}_{2}$. Hence $f=vf_{1}+(1+v)f_{2}=1+vx+(1+v)x^{3}+(1+v)x^{6}$, $\sigma(h)=v+vx^{3}+(1+v)x^{5}+x^{6}$. $C=\langle L(f)\rangle$ is a linear code over $R$ and $\Phi(C)$ is a reversible DNA code. Now we consider the generator matrix of $C$.
\begin{equation}
 \begin{pmatrix}
                    E_{0}   \\
                    F_{0}  \\
                    E_{1}  \\
                    F_{1}  \\
                    E_{2}   \\
                    F_{2}
\end{pmatrix}
= \begin{pmatrix}
                     1 & v & 0 & 1+v & 0 & 0 & 1+v & 0  & 0   \\
                      v & 0 & 0 & v & 0 & 1+v & 1 & 0 &  0  \\
                     0 & 1 & v & 0 & 1+v & 0 & 0 & 1+v & 0   \\
                     0 & v & 0 & 0 & v & 0 & 1+v & 1 & 0  \\
                     0 & 0 & 1 & v & 0 & 1+v & 0 & 0 & 1+v   \\
                     0 & 0 & v & 0 & 0 & v & 0 & 1+v & 1
\end{pmatrix}.
\end{equation}

If we take $\alpha_{0}=0$, $\alpha_{1}=1$, $\alpha_{2}=u$, $\beta_{0}=0$, $\beta_{1}=1$ and $\beta_{2}=v$, then $\alpha_{0}E_{0}+\alpha_{1}E_{1}+\alpha_{2}E_{2}+\beta_{0}F_{0}+\beta_{1}F_{1}+\beta_{2}F_{2}=(1+v)x+ux^{2}+uvx^{3}+x^{4}+(u+v+uv)x^{5}+(1+v)x^{6}+vx^{7}+(u+v+uv)x^{8}$ and this corresponds to the codeword $c_{1}=(0,1+v,u,uv,1,u+v+uv,1+v,v,u+v+uv)$. Hence $\Phi(c_{1})=(AGTAGTGATAATTGGAGG)$. Furthermore, $\sigma(\alpha_{0})F_{2}+\sigma(\alpha_{1})F_{1}+\sigma(\alpha_{2})F_{0}+\sigma(\beta_{0})E_{2}+\sigma(\beta_{1})E_{1}+\sigma(\beta_{2})E_{0}=1+v+uv+(1+v)x+vx^{2}+(1+v+uv)x^{3}+x^{4}+(u+uv)x^{5}+ux^{6}+vx^{7}$ corresponds to the codeword $c_{2}=(1+v+uv,1+v,v,1+v+uv,1,u+uv,u,v,0)$ and thus $\Phi(c_{2})=(GGAGGTTAATAGTGATGA)$. Therefore, $(\Phi(c_{1}))^{r}=\Phi(c_{2})$.\\

\noindent\textbf{Corollary 4.4} \emph{Let $C=vC_{1}\oplus(1+v)C_{2}$ is a cyclic code of arbitrary lengths $n$ over $R$, $C_{1}$ and $C_{2}$ are reversible and $C=\langle L(f)\rangle$ be a linear code over $R$. If $(\overline{0},\overline{0},\cdots,\overline{0})\in C$, then $\Phi(C)$ gives a reversible-complement DNA code.}\\

\noindent\textbf{Proof.} It follows from Theorem 3.8, Theorem 4.2 and Theorem 3.13 immediately.\qed\\

\noindent\textbf{Example 4.5} Let $f_{1}=x+1$ and $f_{2}=x^{6}+x^{5}+x^{4}+x^{3}+x^{2}+x+1$ where both divide $x^{7}-1$ over $\mathbb{F}_{2}$. Hence $C=\langle vf_{1}+(1+v)f_{2}\rangle_{\sigma}=\langle1+x+(1+v)x^{2}+(1+v)x^{3}+(1+v)x^{4}+(1+v)x^{5}+(1+v)x^{6}\rangle_{\sigma}$, is a $\sigma$-linear code over $R$ and  $\Phi(C)$ is a reversible-complement DNA code since
$(\overline{0},\overline{0},\cdots,\overline{0})\in C$.\\

\noindent\textbf{Corollary 4.6} \emph{Let $C=vC_{1}\oplus(1+v)C_{2}$ is a cyclic code of arbitrary lengths $n$ over $R$, $C_{1}$ and $C_{2}$ are reversible and $C=\langle L(f)\rangle$ be a linear code over $R$ and $\Phi(C)$ be a reversible  DNA code. If $(\overline{0},\overline{0},\cdots,\overline{0})$ is added to generator set $L(f)$, then $\Phi(C)$ is a reversible-complement DNA code.}\\

\noindent\textbf{Theorem 4.7} \emph{Let $C_{1}$ is reversible and $f=vf_{1}+(1+v)f_{1}$ over $R$. Then, $C=(f)$ is a reversible cyclic code over $R$ and $\Phi(C)$ is a reversible DNA code. If $x-1\nmid f$, then $\Phi(C)$ is a reversible-complement DNA code.}\\

\noindent\textbf{Proof.} Let the dimension of the code $C$ be $k$. Suppose that the linear cyclic code $C$ has generator matrix with rows $f,xf,\cdots,x^{k-1}f$. If we use the  $\sigma$-set $L(f)$, then we observe that\\
\begin{equation}
(\Phi(\sum\limits_{i}\alpha_{i}x^{i}f))^{r}=\Phi(\sum\limits_{i}\sigma(\alpha_{i})x^{k-1-i}f),
\end{equation}
where $\alpha\in R$ and $0\leq i\leq k-1$ since $f=vf_{1}+(1+v)f_{2}$ and coefficients of $f$ are solely from $R_{1}$ which proves the reversibility in DNA. Thus, we can use the generator matrix of a linear cyclic code or the $\sigma$-set $L(f)$ since $\sigma$ does not effect the coefficients. If $x-1\nmid f$, then $C$ contains $1+x+\cdots+x^{n-1}$. Therefore, $\Phi(C)$ gives a reversible-complement DNA code by Corollary 4.5.\qed\\

The following results come from [12]. Let $R$ be a finite QF ring. As a ring, $R$ admits a decomposition $R=\oplus_{\alpha\in\Delta}Re_{\alpha}$ where $e_{\alpha}$ are central orthogonal idempotents, with $1_{R}=\sum_{\alpha\in\Delta}e_{\alpha}$. Then $R_{\alpha}~:=~Re_{\alpha}$ is also a QF ring for each $\alpha\in\Delta$. If $C$ is a right (resp., left) linear code of length $n$ over $R$, then $C_{\alpha}~:=Ce_{\alpha}$ (resp., $C_{\alpha}~:=e_{\alpha}C$) is a right (resp., left) linear code of length $n$ over $R_{\alpha}$.\\

 \noindent\textbf{Lemma 4.8} [12] \emph{If $C$ is a right (left) linear code of length $n$ over $R$, then\\
(1) $k(C)=max_{\alpha\in\Delta}\{k(C_{\alpha})\}$,\\
(2) $d(C)=max_{\alpha\in\Delta}\{d(C_{\alpha})\}$.}\\

\noindent\textbf{Theorem 4.9} [12] \emph{Let $R=\oplus_{\alpha\in\Delta}R_{\alpha}$ be a finite quasi-Frobenius ring such that $R_{\alpha}$ is a local ring for all $\alpha\in\Delta$ and let $q_{\alpha}$ be the prime power such that $|R_{\alpha}/J(R_{\alpha})|=q_{\alpha}$ for each $\alpha\in\Delta$. If $C$ is a right (left) linear code of length $n$ over $R$, then\\
\begin{equation}
n\geq~\sum\limits_{i}\limits^{k(C)-1}~\lceil d(C)/q^{i}\rceil,
\end{equation}
where $q~:=max_{\alpha\in\Lambda}\{q_{\alpha}\}$.}\\

\dse{5~~The $GC$ weight} As we all known, a DNA code with the same $GC$ weight (content) in every codeword ensures that the codewords have similar thermodynamic characteristics (i.e. melting temperature and hybridization energy). In this section, we will study the $GC$ weight over $R$ by the image of Gray map.\\

In order to study the $GC$ weight over $R$, we give the following lemma first which received from [14] easily.\\

\noindent\textbf{Lemma 5.1} [14] \emph{Let $C^{'}$ be a cyclic code over $R_{1}$. Then there are unique polynomial $g,a,p$ in $\mathbb{F}_{2}[x]$, s.t. $C^{'}=(g+up,ua)$, where $a\mid g\mid x^{n}-1$ and $degp<dega$.}\\

\noindent\textbf{Lemma 5.2} [14] \emph{Let $C^{'}$ be a cyclic code over $R_{1}$. If $(n,2)=1$, then $C^{'}=(g,ua)=(g+ua)$,\\
(1) If $a=g$, we have $C^{'}=(g)$. It is a free-module with rank of $n-degg$ and a set of $\mathbb{F}_{2}$-basis \\
is $\{g,xg,\cdots,x^{n-degg-1}g,ug,uxg,\cdots,ux^{n-degg-1}g\}$;\\
(2) If $a\neq g$, then $C^{'}$ is not a free-module which rank is $n-dega$. A set of $\mathbb{F}_{2}$-basis is\\
$\{g,xg,\cdots,x^{n-degg-1}g,ug,uxg,\cdots,ux^{n-degg-1}g,ua,uxa,\cdots,ux^{degg-dega-1}a\}$.}\\

\noindent\textbf{Lemma 5.3} [14] \emph{Let $C^{'}$ be a cyclic code over $R_{1}$. If $(n,2)\neq1$, then\\
(1) If $a=g$, we have $C^{'}=(g+up)$. It is a free-module with rank of $n-degg$ and a set of\\
$\mathbb{F}_{2}$-basis is $\{g+up,x(g+up),\cdots,x^{n-degg-1}(g+up),ug,uxg,\cdots,ux^{n-degg-1}g\}$;\\
(2) If $a\neq g$, then $C^{'}$ is not a free-module which rank is $n-dega$. A set of $\mathbb{F}_{2}$-basis is\\
$\{g+up,x(g+up),\cdots,x^{n-degg-1}(g+up),ug,uxg,\cdots,ux^{n-degg-1}g,ua,xua,\cdots,x^{degg-dega-1}ua\}$.}\\

Using the lemmas above and the structure of $C$ we have already received, we can get the following Theorem immediately.\\

\noindent\textbf{Theorem 5.4} \emph{Let $C=vC_{1}\oplus(1+v)C_{2}$ be a cyclic code of arbitrary lengths $n$ over $R$, where $C_{1}$ and $C_{2}$ are both cyclic codes over $R_{1}$. Then $C$ has a minimal generating set $\Gamma=v\Pi+(1+v)\Omega$, where $\Pi,\Omega$ are the minimal generating set of $C_{1}$ and $C_{2}$, respectively.}\\

Now we have already had the minimal generating set of $C$, so we can study its Gray images.

On account of the minimal generating set of $C$ is $\Gamma=v\Pi+(1+v)\Omega$ and the Gray map from $R$ to $R^{2}_{1}$ is defined as $\xi(a+bv)=(a,a+b)$, we can get the Gray images of the minimal generating set of $C$ is  $\Phi(\Gamma)=x^{n}\Pi+\Omega$, where $\Pi,\Omega$ are the minimal generating set of $C_{1}$ and $C_{2}$, respectively.

If we can prove $\zeta$ is a linear transformation, then the $GC$ weight over $R$ is given by the Hamming weight of the $u\Phi(\Gamma)=u(x^{n}\Pi+\Omega)$.

  For any $x=a_{1}+b_{1}v$, $y=a_{2}+b_{2}v\in C$,
  \begin{equation}
\xi(x+y)=(a_{1}+a_{2})+(a_{1}+a_{2}+b_{1}+b_{2})v=\xi(x)+\xi(y).
\end{equation}\\
\noindent\textbf{Theorem 5.5} \emph{Let $C=vC_{1}\oplus(1+v)C_{2}$ be a cyclic code of arbitrary lengths $n$ over $R$, $C_{1}=(g_{1}+up_{1},ua_{1})$, $C_{2}=(g_{2}+up_{2},ua_{2})$, where $a_{1}\mid g_{1}\mid x^{n}-1$, $a_{2}\mid g_{2}\mid x^{n}-1$, $degp_{1}<dega_{1}$ and $degp_{2}<dega_{2}$. Then the $GC$ weight over $R$ is given by the Hamming weight enumerator of the
  \begin{equation}
\Lambda=x^{n}\{g_{1},xg_{1},\cdots,x^{n-degg_{1}-1}g_{1}\}+\{g_{2},xg_{2},\cdots,x^{n-degg_{2}-1}g_{2}\}.
\end{equation}}

\noindent\textbf{Proof.} The $GC$ content is obtained by multiplying the Gray images of the minimal generating set of $C$ by $u$, and from Theorem 5.4 we have
  \begin{equation}
u\Phi(\Gamma)=ux^{n}\{g_{1},xg_{1},\cdots,x^{n-degg_{1}-1}g_{1}\}+u\{g_{2},xg_{2},\cdots,x^{n-degg_{2}-1}g_{2}\}.
\end{equation}
Hence the $GC$ content is given by the Hamming weight of the
  \begin{equation}
\Lambda=x^{n}\{g_{1},xg_{1},\cdots,x^{n-degg_{1}-1}g_{1}\}+\{g_{2},xg_{2},\cdots,x^{n-degg_{2}-1}g_{2}\}.
\end{equation}\qed\\

At the end of this section, we give some examples to illustrate the main work in this paper.\\

\noindent\textbf{Example 5.6} Let $x^{3}-1=(x+1)(x^{2}+x+1)=g_{1}g_{2}\in F_{2}[x]$. Let $C_{1}=C_{2}=(g,ua)$ be a cyclic code of length $3$ over $R_{1}$, where $g=g_{2},~a=g_{2}$. The image of $C$ under the Gray map $\Phi$ is a DNA code of length $6$. This code has $16$ codewords which are listed in the table 2.\\

\begin{table}[htbp]
{\small
\begin{center}
{\small{\bf Table 2 }~~All $16$ codewords of $C$}\\

\begin{tabular}{l l l l}\\
\hline
 $AAAAAA$ & $GGGGGG$ & $TTTTTT$ & $CCCCCC$ \\
 $AAAGGG$ & $GGGAAA$ & $TTTCCC$ & $CCCTTT$ \\
 $AAATTT$ & $GGGCCC$ & $TTTAAA$ & $CCCGGG$ \\
 $AAACCC$ & $GGGTTT$ & $TTTGGG$ & $CCCAAA$ \\
\hline
\end{tabular}
\end{center}}
\end{table}

In the following example, we obtain some optimal codes over $R$ where $f_{1}=f_{2}$ which satisfy the max Griesmer bound given by Leo et al. [17].\\

\noindent\textbf{Example 5.7} Let $f_{1}=1+x^{2}+x^{4}+x^{6}=f_{2}$ be a self-reciprocal polynomial where $f_{1}\mid x^{8}-1$ over $R$. $C=\langle L(f)\rangle$ is a cyclic linear code over $R$ that attains the maximum Griesmer bound on $R$ with parameters [8,2,4]. Also $\Phi(C)$ is a reversible DNA code which is not complement because $(x+1)\mid f_{1}$. We assign the DNA bases $A,T,G,C$ to $0,1,2$ and $3$, respectively and a DNA string is converted to quaternary number system and then to the decimal system to save some space in table 3. For instance, $859024042$ represents $ACACACACGGGGGGGG$.\\

\dse{6~~Conclusion}
Algebraic structure of codes have already acquired over the non-chain ring $R$ with $16$ elements. The DNA codes over $R$ are studied which are obtained by using a special auto-morphism and properties of cyclic codes. We introduced these codes correspond to reversible and reversible-complement DNA codes with DNA double pairs by means of a special DNA corresponding table. Finally, the $GC$ weight over $R$ is studied by using the image of Gray map.

\begin{table}[htbp]
{\small
\begin{center}
{\small{\bf Table 3 }~~DNA correspondence of $C=\langle L(f)\rangle$ explained in Example 5.7}\\
\begin{tabular}{l l l l l l l}\\
\hline
$0$ & $8738$ & $4369$ & $13107$ & $572662306$ & $572653568$ & $572666675$\\
$572657937$ & $286331153$ & $286339891$ & $286326784$ & $286335522$ & $858993459$ & $858984721$  \\
$858989090$ & $858980388$ & $34952$ & $43690$ & $39321$ & $48059$ & $572697258$  \\
$572688520$ & $572701627$ & $572692889$ & $286366105$ & $286374843$ & $286361736$ & $286390474$ \\
$859028411$ & $859019673$ & $859024042$ & $859015304$ & $17476$ & $26214$ & $21845$  \\
$30583$ & $572679782$ & $572671044$ & $572684151$ & $572675413$ & $286348629$ & $286357367$ \\
$286344260$ & $286352998$ & $859010935$ & $859002197$ & $859006566$ & $858997828$ & $52428$ \\
$61166$ & $56797$ & $65535$ & $572714734$ & $572705996$ & $572719103$ & $572710365$ \\
$286383581$ & $286392319$ & $286379212$ & $286387950$ & $859045887$ & $859037149$ & $859041518$  \\
$859032780$ & $2290649224$ & $2290657962$ & $2290653593$ & $2290662331$ & $2863311530$ & $2863302792$ \\
$2863315899$ & $2863307161$ & $2576980377$ & $2576989115$ & $2576976008$ & $2576984746$ & $3149642683$ \\
$3149633945$ & $3149638314$ & $3149629576$ & $2290614272$ & $2290623010$ & $2290618641$ & $2290627379$ \\
$2863276578$ & $2863267840$ & $2863280947$ & $2863272209$ & $2576945425$ & $2576954163$ & $2576941056$  \\
$2576949794$ & $3149607731$ & $3149598993$ & $3149603362$ & $3149594624$ & $2290666700$ & $2290675438$ \\
$2290671069$ & $2290679807$ & $2863329006$ & $2863320268$ & $2863333375$ & $2863324637$ & $2576997853$ \\
$2577006591$ & $2576993484$ & $2577002222$ & $3149660159$ & $3149651421$ & $3149655790$ & $3149647052$ \\
$2290631748$ & $2290640486$ & $2290636117$ & $2290644855$ & $2863294054$ & $2863285316$ & $2863298423$  \\
$2863289685$ & $2576962901$ & $2576971639$ & $2576958532$ & $2576967270$ & $3149625207$ & $3149616469$ \\
$3149620838$ & $3149612100$ & $1145324612$ & $1145333350$ & $1145328981$ & $1145337719$ & $1717986918$ \\
$1717978180$ & $1717991287$ & $1717982549$ & $1431655765$ & $1431664503$ & $1431651396$ & $1431660134$  \\
$2004318071$ & $2004309333$ & $2004313702$ & $2004304964$ & $1145359564$ & $1145368302$ & $1145363933$ \\
$1145372671$ & $1718021870$ & $1718013132$ & $1718026239$ & $1718017501$ & $1431690717$ & $1431699455$ \\
$1431686348$ & $1431695086$ & $2004353023$ & $2004344285$ & $2004348654$ & $2004339916$ & $1145307136$  \\
$1145315874$ & $1145311505$ & $1145320243$ & $1717969442$ & $1717960704$ & $1717973811$ & $1717965073$ \\
$1431638289$ & $1431647027$ & $1431633920$ & $1431642658$ & $2004300595$ & $2004291857$ & $2004296226$ \\
$2004287488$ & $1145342088$ & $1145350826$ & $1145346457$ & $1145355195$ & $1718004394$ & $1717995656$  \\
$1718008763$ & $1718000025$ & $1431673241$ & $1431681979$ & $1431668872$ & $1431677610$ & $2004335547$ \\
$2004326809$ & $2004331178$ & $2004322440$ & $3435973836$ & $3435982574$ & $3435978205$ & $3435986943$ \\
$4008636142$ & $4008627404$ & $4008640511$ & $4008631773$ & $3722304989$ & $3722313727$ & $3722300620$  \\
$3722309358$ & $4294967295$ & $4294958557$ & $4294962926$ & $4294954188$ & $3435938884$ & $3435947622$ \\
$3435943253$ & $3435951991$ & $4008601190$ & $4008592452$ & $4008605559$ & $4008596821$ & $3722270037$ \\
$3722278775$ & $3722265668$ & $3722274406$ & $4294932343$ & $4294923605$ & $4294927974$ & $4294919236$  \\
$3435956360$ & $3435965098$ & $3435960729$ & $3435969467$ & $4008618666$ & $4008609928$ & $4008623035$ \\
$4008614297$ & $3722287513$ & $3722296251$ & $3722283144$ & $3722291882$ & $4294949819$ & $4294941081$ \\
$4294945450$ & $4294936712$ & $3435921408$ & $3435930146$ & $3435925777$ & $3435934515$ & $4008583714$  \\
$4008574976$ & $4008588083$ & $4008579345$ & $3722252561$ & $3722261299$ & $3722248192$ & $3722256930$ \\
$4294914867$ & $4294906129$ & $4294910498$ & $4294901760$ \\
\hline
\end{tabular}
\end{center}}
\end{table}

\end{document}